\documentclass{article}

\begin{document}

\newcommand{\half}{\frac{1}{2}}

\thispagestyle{empty}

\begin{center}
 {\Large
    Instantons on General Noncommutative $\bf{R}^4$
 }
\end{center}

\vspace*{2cm}
\begin{center}
 \noindent
 {\large Yu Tian}
 \vspace{5mm}
 \noindent
 \hspace{0.7cm} \parbox{142mm}{\it
School of Physics, Peking University, Beijing 100871, China
\\
E-mail: {\tt phytian@yahoo.com}
 }
\end{center}
 \vspace{5mm}
\begin{center}
 \noindent
 {\large Chuan-Jie Zhu}
 \vspace{5mm}
 \noindent
 \hspace{0.7cm} \parbox{142mm}{\it
Institute of Mathematics, Henan University, Kaifeng 475001
\\
and
\\
Institute for Theoretical Physics, Chinese Academy of Sciences
\\
P. O. Box 2735, Beijing 100080
\\
E-mail: {\tt zhucj@itp.ac.cn}
 }
\end{center}

\vspace{2cm}
\hfill{\bf Abstract\ \ }\hfill\ \\

We study the $U(1)$ and $U(2)$ instanton solutions of gauge theory
on general noncommutative $\bf{R}^4$. In all  cases considered we
obtain explicit results for the projection operators. In some
cases we computed numerically the instanton charge and found that
it is an integer, independent of the noncommutative parameters
$\theta_{1,2}$.

\newpage

\section{Introduction}

The study of exact solutions in field theory is a very important
subject. It is often the first step to learn something
nonperturbatively about a given theory.

In recent years the study of noncommutative field theory becomes
an active research area, mostly due to its relevance with string
theory \cite{WittenSeiberg}. Perturbative analysis of
noncommutative field theory reveals an interesting inter-relation
between infrared and ultraviolet divergences \cite{Seiberg}. This
is due to the presence of a short distance like cutoff in the
noncommutative space on which the theory is formulated. Other
peculiar features include the exact soliton solution in pure
scalar field theory \cite{GSM} and a vast of other interesting
exact solutions in noncommutative field theories \cite{Harvey}.

Instantons are exact solutions in gauge field theory. These
solutions are also interesting in mathematics \cite{Donaldson}.
Recently, noncommutative instantons \cite{Schwarz} become one of
the great interests in theoretical physics.

In this paper we will study instantons in noncommutative gauge
theory. In particular we focus our study on how the involved
quantities vary with the noncommutative parameters $\theta^{mn}$.
We note that the usual treatment of setting $\theta_1=\pm\theta_2$
(see below for our notations) by rescaling the coordinates $x^m$
on general noncommutative $\bf{R}^4$ (or in brief, $\bf{R}_{\rm
NC}^4$) is not allowed because it will change the metric and so
the (anti-)self-dual equations. In this paper we will keep this
arbitrariness and study in detail the various explicit solutions
in noncommutative $U(1)$ and $U(2)$ gauge theory. We note that
these solutions are studied for the case $\theta_1=\pm\theta_2$ in
\cite{Papera,Paperb,Paperc,Paperd,Hamanaka}.

\section{$\bf{R}_{\rm  NC}^4$ and the (anti-)self-dual equations}
First let us recall briefly the noncommutative $\bf{R}^4$ and set
our notations\footnote{For general reviews on noncommutative
geometry and field theory, see, for example, \cite{Paperc,
Reviewa, Harvey, Reviewb, Reviewc}.}. For a general noncommutative
$\bf{R}^4$ we mean a space with (operator) coordinates $x^m$,
$m=1, \cdots, 4$, which satisfy the following relations:
\begin{equation}
[x^m,x^n] = i \theta^{mn},
\end{equation}
where $\theta^{mn}$ are real constants. If we assume the standard
(Euclidean) metric for the noncommutative $\bf{R}^4$, we
can use the orthogonal transformation with positive determinant to
change $\theta^{mn}$ into the following standard form:
\begin{equation}
\label{theta} (\theta^{mn})=\left(\begin{array}{cccc} 0 &
\theta^{12} & 0 & 0 \\ -\theta^{12}& 0 & 0 & 0 \\ 0 & 0 & 0
&\theta^{34}\\ 0 & 0 & -\theta^{34}& 0 \end{array}\right),
\end{equation}
where $\theta^{12}>0$ and $\theta^{12}+\theta^{34}\geq 0$. By
using this form of $\theta^{mn}$, the only non-vanishing
commutators are as follows:
\begin{equation}
[x^1,x^2] = i \theta^{12}, \qquad [x^3,x^4] = i \theta^{34},
\end{equation}
and other twos obtained by using the anti-symmetric property of
the commutators. Introducing complex coordinates:
\begin{equation}
\label{complex}
\begin{array}{rl}
z_1 = x^2 + i x^1 , & \bar{z}_1 = x^2 - i x^1, \\
z_2 = x^4 + i x^3 , & \bar{z}_2 = x^4 - i x^3,
\end{array}
\end{equation}
the non-vanishing commutation relations are
\begin{equation}
\label{commutator of z}
[\bar z_1,z_1]=2\theta^{12}\equiv\theta_1,\quad [\bar z_2,z_2]
=2\theta^{34}\equiv\theta_2.
\end{equation}

By a noncommutative gauge field $A_m$ we mean an operator valued
field. The (anti-hermitian) field strength $F_{mn}$ is defined
similarly as in the commutative case:
\begin{equation}
\label{F by A}
F_{mn}=\hat\partial_{[m} A_{n]} + A_{[m} A_{n]}
\equiv \hat\partial_m A_n - \hat\partial_n A_m + [A_m, A_n],
\end{equation}
where the derivative operator $\hat\partial_m$ is defined as follows:
\begin{equation}
\hat\partial_m f \equiv - i \theta_{mn} [x^n, f],
\end{equation}
where $\theta_{mn}$ is the inverse of $\theta^{mn}$.
For our standard form (\ref{theta}) of $\theta^{mn}$ we have
\begin{equation}
\hat\partial_1 A = {i \over \theta^{12}} [x^2,A],
\qquad \hat\partial_2 A = - {i \over \theta^{12}} [x^1,A],
\end{equation}
which can be expressed by the complex coordinates (\ref{complex}) as
follows:
\begin{equation}
\partial_1 A \equiv \hat
\partial_{z_1} A = {1\over \theta_1} [\bar z_1, A],
\qquad \bar\partial_1 A \equiv \hat\partial_{\bar z_1} A
= - {1\over \theta_1} [z_1, A],
\end{equation}
and similar relations for $x^{3,4}$ and $z_2,\bar{z}_2$.

For a general metric $g_{mn}$ the instanton equations are
\begin{equation}
\label{instanton}
F_{mn}=\pm\frac{\epsilon^{pqrs}}{2\sqrt{g}}g_{mp}g_{nq}F_{rs},
\end{equation}
and the solutions are known as self-dual (SD, for ``+'' sign) and
anti-self-dual (ASD, for ``$-$'' sign) instantons. Here
$\epsilon^{pqrs}$ is the totally anti-symmetric tensor
($\epsilon^{1234}=1$ etc.) and $g$ is the metric. We will
take the standard metric $g_{mn}=\delta_{mn}$ and take the
noncommutative parameters $\theta_{1,2}$ as free parameters. We
also note that the notions of self-dual and anti-self-dual are
interchanged by a parity transformation. A parity transformation
also changes the sign of $\theta^{mn}$. In the following
discussion we will consider only the ASD instantons. So we should
not restrict $\theta_2$ to be positive.

\section{Instantons in Noncommutative Gauge Theory}

\subsection{ADHM construction for ordinary gauge theory}

For ordinary gauge theory all the (ASD) instanton solutions are
obtained by ADHM (Atiyah-Drinfeld-Hitchin-Manin) construction
\cite{ADHM}. In this construction we introduce the following
ingredients (for $U(N)$ gauge theory with instanton number $k$):
\begin{itemize}
\item complex vector spaces $V$ and $W$ of dimensions $k$ and $N$,
\item $k\times k$ matrix $B_{1,2}$, $k\times N$ matrix $I$ and
$N\times k$ matrix $J$,
\item the following quantities:
\begin{eqnarray}
\label{ADHM1}
\mu_r & = & [B_1, B_1^\dagger] + [B_2, B_2^\dagger] + I \,
I^\dagger -J^\dagger J, \\
\label{ADHM2}
\mu_c & = & [B_1,B_2] + I\, J.
\end{eqnarray}
\end{itemize}
The claim of ADHM is as follows:
\begin{itemize}
\item Given $B_{1,2}$, $I$ and $J$ such that $\mu_r=\mu_c=0$, an
ASD gauge field can be constructed;
\item All ASD gauge fields can be obtained in this way.
\end{itemize}

It is convenient to introduce a quaternionic notation\footnote{We
follow closely the notation of \cite{Paperd}.} for the
4-dimensional Euclidean space-time indices:
\begin{equation}
x\equiv x^n\sigma_n,\qquad\bar{x}\equiv x^n\bar\sigma_n,
\end{equation}
where $\sigma_n=(i\vec{\tau},1)$ and $\tau^c$, $c=1,2,3$ are the
three Pauli matrices, and the conjugate matrices
$\bar\sigma_n=\sigma_n^\dag=(-i\vec{\tau},1)$. In terms of the
complex coordinates (\ref{complex}) we have
\begin{equation}
(x_{\alpha\dot\alpha})=\left(\begin{array}{cc} z_2 & z_1 \\ -
\bar{z}_1 & \bar{z}_2 \end{array}\right), \qquad
(\bar{x}^{\dot\alpha\alpha})=\left(\begin{array}{cc} \bar{z}_2 &
- z_1 \\ \bar{z}_1 & z_2 \end{array}\right).
\end{equation}
Then the basic object in the ADHM construction is the $(N+2k)\times
2k$ matrix $\Delta$ which is linear in the space-time coordinates:
\begin{equation}
\label{Delta}
\Delta=a+b\bar{x},
\end{equation}
where the constant matrices
\begin{equation}
a = \left( \begin{array}{cc} I^\dag & J \\ B_2^\dagger & -B_1
\\ B_1^\dagger & B_2 \end{array} \right), \quad
b = \left( \begin{array}{cc} 0 & 0 \\ 1 & 0 \\ 0 & 1 \end{array}
 \right).
\end{equation}

Consider the conjugate operator of $\Delta$:
\begin{equation}
\Delta^\dagger = a^\dag +x b^\dag = \left( \matrix{I & B_2 +
z_2 & B_1 + z_1 \cr J^\dagger & -B_1^\dagger -\bar z_1 &
B_2^\dagger + \bar z_2} \right).
\end{equation}
It is easy to check that the ADHM equations (\ref{ADHM1}) and
(\ref{ADHM2}) are equivalent to the so-called factorization
condition:
\begin{equation}
\label{factorize}
\Delta^\dagger\Delta=\left(\matrix{f^{-1} & 0 \cr 0 & f^{-1}}\right),
\end{equation}
where $f(x)$ is a  $k\times k$ hermitian matrix. From the above
condition we can construct a hermitian projection operator $P$ as
follows:
\begin{eqnarray}
P&=&\Delta f\Delta^\dag, \cr
P^2&=&\Delta f f^{-1}f\Delta^\dag=P.
\end{eqnarray}

Obviously, the null-space of $\Delta^\dagger(x)$ is of $N$
dimension for generic $x$. The basis vector for this null-space
can be assembled into an $(N+2k)\times N$ matrix $U(x)$:
\begin{equation}
\Delta^\dag U=0,
\end{equation}
which can be chosen to satisfy the following orth-normalization
condition:
\begin{equation}
U^\dag U=1.
\end{equation}
The above orth-normalization condition guarantees that $UU^\dag$
is also a hermitian projection operator. Now it can be proved that
the completeness relation
\begin{equation}
\label{complete}
P+UU^\dag=1
\end{equation}
holds if $U$ contains the whole null-space of $\Delta^\dagger$. In
other words, this completeness relation requires that $U$ consists
of all the zero modes of $\Delta^\dagger$. The proof is sketched
as follows: the two projection operator $P$ and $UU^\dag$ are
orthogonal to each other, and so $1-P-UU^\dag$ is also a hermitian
projection operator. Now this can always be written as the form
$VV^\dag$; then $V$ must consist of some zero modes of
$\Delta^\dagger$ other than those in $U$ because $\Delta$ and $f$
are both of maximum rank (this notion is ambiguous in the
infinite-dimensional case, but some other notions can be used
instead) and $PVV^\dag=0$. This conclusion is in conflict with the
assumption that $U$ contains all the zero modes of
$\Delta^\dagger$.

The (anti-hermitian) gauge potential is constructed from $U$ by
the following formula:
\begin{equation}
A_m= U^\dag\partial_m U.
\end{equation}
Substituting this expression into (\ref{F by A}), we get the
following field strength:
\begin{eqnarray}
\label{calculate F}
F_{mn}&=&\partial_{[m}(U^\dag\partial_{n]}U)
+(U^\dag\partial_{[m}U)(U^\dag\partial_{n]}U)
=\partial_{[m}U^\dag(1-UU^\dag)\partial_{n]}U\nonumber\\
&=&\partial_{[m}U^\dag\Delta f\Delta^\dag\partial_{n]}U
=U^\dag\partial_{[m}\Delta f\partial_{n]}\Delta^\dag U
=U^\dag b\bar\sigma_{[m}\sigma_{n]}f b^\dag U\nonumber\\
&=& 2i\bar\eta^c_{mn}U^\dag b(\tau^c f)b^\dag U.
\end{eqnarray}
Here $\bar\eta^a_{ij}$ is the standard 't Hooft $\eta$-symbol,
which is anti-self-dual:
\begin{equation}
\half\epsilon_{ijkl}\bar\eta^a_{kl}=-\bar\eta^a_{ij}.
\end{equation}

\subsection{Noncommutative ADHM construction}

The above construction has been extended to noncommutative gauge
theory \cite{Schwarz}. We recall this construction briefly here.
By introducing the same data as above but considering  the $z_i$'s
as noncommutative we see that the factorization condition
(\ref{factorize}) still gives $\mu_c=0$, but $\mu_r$ no longer
vanishes. It is easy to check that the following relation holds:
\begin{equation} \mu_r=\zeta\equiv\theta_1+\theta_2.
\end{equation}
In this case the two ADHM equations (\ref{ADHM1}) and
(\ref{ADHM2}) can be combined into one \cite{Paperd}:
\begin{equation}
\label{ADHM} \tau^{c\dot\alpha}{}_{\dot\beta}(\bar a^{\dot\beta}
a_{\dot\alpha})_{ij}=\delta_{ij}\delta^{c3}\zeta.
\end{equation}

As studied mathematically by various people (see, for example, the
lectures by H. Nakajima \cite{Nakajima}), the moduli space of the
noncommutative instantons is better behaved than their commutative
counterpart. In the noncommutative case the operator
$\Delta^\dagger\Delta$ always has maximum rank, i.e., it has no
zero modes (see \cite{Reviewa}).

Though there is no much difference between the noncommutative ADHM
construction and the commutative one, we should study the
noncommutative case in more detail. One important problem is that
the instanton charge is not evidently integer. In order to study
the instanton solution precisely, we use a Fock space
representation as follows ($n_1, n_2\geq 0$):
\begin{eqnarray}
z_1|n_1,n_2\rangle & = &
\sqrt{\theta_1}\sqrt{n_1+1}|n_1+1,n_2\rangle,
\\
\quad\bar z_1|n_1,n_2\rangle & =&
\sqrt{\theta_1}\sqrt{n_1}|n_1-1,n_2 \rangle,
\end{eqnarray}
by using the commutation relation (\ref{commutator of z}). Similar
expressions for $z_2$ and $\bar z_2$ also apply (but paying a
little attention to the sign of $\theta_2$ which is not restricted
to be positive). In this representation the $z_i$'s are
infinite-dimensional matrices, and so are the operator $\Delta$,
$\Delta^\dag$ etc. Because of infinite dimensions are involved we
can not determine the dimension of null-space of $\Delta^\dag$
straightforwardly from the difference of the numbers of its rows
and columns. But it turns out that $\Delta^\dag$ also has
infinite number of zero modes, and they can be arranged into an
$(N+2k)\times N$ matrix with entries from the (noncommutative)
algebra generated by the coordinates, which resembles the
commutative case.

In the following sections, we will study in full detail the ASD
1-instanton and 2-instanton solutions of $U(1)$ theory and
1-instanton solutions of $U(2)$ theory on general $\bf{R}_{\rm
NC}^4$. For each of them, the two distinct cases, $\theta_2>0$ and
$\theta_2<0$, are considered separately and all the details of the
zero modes are worked out. The instanton charge is numerically
computed to be integer in the $U(1)$ 1-instanton cases.

\section{$U(1)$ 1-instanton solution}

In this case, the ADHM matrix (\ref{Delta}) which satisfies
(\ref{ADHM}) is given by
\begin{equation}
\Delta=\left(\begin{array}{cc} \sqrt{\zeta}& 0 \\ \bar z_2 & -z_1
\\ \bar z_1 & z_2 \end{array}\right),
\quad\Delta^\dagger=\left(\begin{array}{ccc} \sqrt{\zeta}& z_2 &
z_1
\\ 0 & -\bar z_1 & \bar z_2 \end{array}\right),
\end{equation}
($\zeta=2\theta^{12}+2\theta^{34}\geq 0$ for our assumption) when
the center of mass collective coordinates set to zero. It is
straightforward to obtain
\begin{equation}
f=(Z_1+Z_2+\zeta)^{-1}
\end{equation}
where $Z_1\equiv z_1\bar z_1$ and $Z_2\equiv z_2\bar z_2$.

Now we construct the matrix $U$. It is easy to find a general
solution $U_0$:
\begin{equation}
\label{U_0}
U_0=\left(\begin{array}{c} Z_1+Z_2 \\ -\sqrt{\zeta}\bar z_2
\\ -\sqrt{\zeta}\bar z_1 \end{array}\right),\quad\Delta^\dagger
 U_0=0.
\end{equation}
But the problem is that this $U_0$ is either over-complete or
incomplete for each $\theta_2$ cases. We will solve this problem
in the following.

\subsection{$\theta_2>0$ case}
In this case $\bar z_1$ and $\bar z_2$ are annihilation operators
and  $U_0$ above obviously annihilates the vacuum, i.e.
\begin{equation}
U_0|0,0\rangle=0.
\end{equation}
In other words, as an infinite-dimensional matrix in the Fock space
representation, $U_0$ has a redundant column with all its elements
vanishing. This column can be removed by a shift operator $u^\dag$
which projects out the vacuum:
\begin{equation}
u^\dag u=p\equiv 1-p_0,\quad uu^\dag=1,
\end{equation}
where $p_0=|0,0\rangle\langle 0,0|$. By using this projection the
normalized $U$ which satisfies the completeness relation
(\ref{complete}) can be obtained as follows:
\begin{equation}
U=\tilde U_0\beta,\quad U^\dag U=1,
\end{equation}
where $\tilde U_0=U_0u^\dag$ is the exact set of all zero modes
and
\begin{equation}
\beta=(\tilde U_0^\dag \tilde U_0)^{-1/2}=[u(Z_1+Z_2)(Z_1+Z_2
+\zeta)u^\dag]^{-1/2}\equiv u\beta_p u^\dag
\end{equation}
is a normalization factor.

Now we compute the field strength. It is given  by
(\ref{calculate F}) which turns out to be as follows:
\begin{equation}
\label{F}
F_{mn}=2i\bar\eta^c_{mn}U^\dag b(\tau^c f)b^\dag U,
\end{equation}
and so can be written as
\begin{equation}
F=\zeta\beta u[(z_2 f\bar z_2-z_1 f\bar z_1)(dz_1 d\bar
z_1-dz_2 d\bar z_2)+2z_1 f\bar z_2 d\bar z_1 dz_2+2z_2 f\bar
z_1 d\bar z_2 dz_1]u^\dag\beta.
\end{equation}
The topological instanton charge is then
\begin{equation}
Q=-\frac{1}{8\pi^2}\int F\wedge F=-\frac{\zeta^2}{\pi^2}\int
dx^4 uTu^\dag=-\zeta^2|\theta_1\theta_2|{\rm Tr}_{\cal H}(uTu^\dag)
\end{equation}
where the expression $T$ is only defined on $p{\cal H}$:
\begin{equation}
\begin{array}{rl}
T&=[\beta_p(z_2 f\bar z_2-z_1 f\bar z_1)\beta_p]^2+2\beta_p z_1
f\bar z_2\beta_p^2 z_2 f\bar z_1\beta_p+2\beta_p z_2 f\bar z_1
\beta_p^2 z_1 f\bar z_2\beta_p\\
&=\frac{1}{(Z_1+Z_2)(Z_1'+Z_2')}\left[(\frac{Z_2}{Z_1'+Z_2}-
\frac{Z_1}{Z_1+Z_2'})^2\frac{1}{(Z_1+Z_2)(Z_1'+Z_2')}\right.
\\
&\quad\left.+\frac{2Z_1 Z_2'}{(Z_1+Z_2')^2(Z_1+Z_2'-\theta_1)
(Z_1+Z_2'+\theta_2)}+\frac{2Z_2 Z_1'}{(Z_1'+Z_2)^2(Z_1'+Z_2-
\theta_2)(Z_1'+Z_2+\theta_1)}\right],
\end{array}
\end{equation}
where $Z_1'$ means $Z_1+\theta_1$, $Z_2'$ means $Z_2+\theta_2$.

Notice that $Z_1$ and $Z_2$ have eigenvalues $n_1\theta_1,n_1 \geq
0$ and $n_2\theta_2,n_2\geq 0$ on ${\cal H}$ respectively, we get
then
\begin{equation}
\label{sp}
{\rm Tr}_{\cal H}(uTu^\dag)=\sum_{n_1+n_2>0}T|_{Z_1=n_1\theta_1,
Z_2=n_2\theta_2}.
\end{equation}
Unlike the familiar $\theta_1=\theta_2$ case, the expression in
(\ref{sp}) seems too complicated to be easily worked out by
analytic method. We didn't try hard to sum them analytically. A
simple numerical calculation should be sufficient to convince us
what is the final result.  For reasonable $\theta_{1,2}$, the
series converge quite fast. For example, for
$\theta_1=1.6,\theta_2=0.4$ we have
\begin{equation}
Q(n_1,n_2\leq 200)|_{\theta_1=1.6,\theta_2=0.4}=-0.999895,
\end{equation}
by using the popular software Mathematica. When $\theta_2$ tends
to 0 (fixing $\theta_1$), the series (\ref{sp}) seems to blow up
and its convergency decreases rapidly. In such cases, we should
properly adjust the range of summation and still we get the
satisfying result $Q=-1$. For example
\begin{equation}
Q(n_1\leq 20,n_2\leq 2000)|_{\theta_1=1.99,\theta_2=0.01}=-0.998137.
\end{equation}
These numerical results strongly convince us that $Q=-1$ should be the
right answer.

\subsection{$\theta_2<0$ case}
In this case $\bar z_1$ is an annihilation operator and $\bar z_2$
is a creation operator. The matrix $U_0$ can be directly
normalized:
\begin{equation}
\tilde{U}=U_0\beta,\quad\tilde{U}^\dag\tilde{U}=1
\end{equation}
where
\begin{equation}
\beta=(U_0^\dag U_0)^{-1/2}=[(Z_1+Z_2)(Z_1+Z_2+\zeta)]^{-1/2}.
\end{equation}
But the problem is that $\tilde{U}$ does not satisfy the
completeness relation. Explicitly we have
\begin{equation} (\Delta
f\Delta^\dag+\tilde{U}\tilde{U}^\dag)\left(
\begin{array}{c} 0 \\ |0,0\rangle \\ 0 \end{array}\right)=0
\neq\left(\begin{array}{c} 0 \\ |0,0\rangle \\ 0 \end{array}
\right).
\end{equation}
So $\tilde{U}$ is not the right answer. In fact, $\tilde{U}$
contains almost all the zero modes of $\Delta^\dag$ except one. We
can simply add an extra column to $\tilde{U}$ to make it complete:
\begin{equation}
U=\left(\begin{array}{c} 0 \\ p_0 \\ 0 \end{array}\right)+\tilde{U}u
\end{equation}
where $u$ is the same shift operator introduced in the last
subsection. It is straightforward to check that the completeness
relation (\ref{complete}) is now satisfied, and so the ASD
instanton solution in the $\theta_2<0$ case can been deduced.

The field strength $F$ is again given by (\ref{F}), but we will
not go on to give the lengthy expression of its explicit form here
and in the following sections. The topological instanton charge is
\begin{equation}
Q=-\frac{1}{8\pi^2}\int F\wedge F=-|\theta_1\theta_2| {\rm
Tr}_{\cal H}T
\end{equation}
where
\begin{equation}
\begin{array}{rl}
T&=[(\sqrt{\zeta}u^\dag\beta z_2-p_0)f(\sqrt{\zeta}\bar
z_2\beta u-p_0)-\zeta u^\dag\beta^2 z_1 f\bar z_1 u]^2\\
&\quad+2\zeta u^\dag\beta^2 z_1 f^2(\zeta\bar z_2\beta^2
z_2+p_0)\bar z_1 u\\
&\quad+2\zeta(\sqrt{\zeta}u^\dag\beta z_2-p_0)f^2\bar z_1
\beta^2 z_1(\sqrt{\zeta}\bar z_2\beta u-p_0),
\end{array}
\end{equation}
and so
\begin{equation}
\begin{array}{rl}
{\rm Tr}_{\cal H}T&=\sum_{n_1\geq 0,n_2\geq 1}\{\zeta^2
\beta^4[Z_2(Z_1'+Z_2)^{-1}-Z_1(Z_1+Z_2')^{-1}]^2\\
&\quad+2\zeta\beta^2 Z_1 Z_2'(Z_1+Z_2')^{-2}(Z_1+Z_2'-
\theta_1)^{-1}(Z_1+Z_2'+\theta_2)^{-1}\\
&\quad+2\zeta\beta^2 Z_2 Z_1'(Z_1'+Z_2)^{-2}(Z_1'+Z_2-
\theta_2)^{-1}(Z_1'+Z_2+\theta_1)^{-1}\}\\
&\quad+\theta_1^{-2}+2\zeta\theta_1^{-2}(\theta_1- \theta_2)^{-1},
\end{array}
\end{equation}
The same numerical calculations give the following results:
\begin{equation}
Q(n_1,n_2\leq 200)|_{\theta_1=2.4,\theta_2=-0.4}=-0.999895,
\end{equation}
\begin{equation}
Q(n_1\leq 20,n_2\leq 2000)|_{\theta_1=2.01,\theta_2=-0.01}
=-0.998141,
\end{equation}
and so we get $Q=-1$ which agrees with the result obtained by the
well-known argument.

\section{$U(1)$ 2-instanton solution}

The moduli space of 2-instanton is much more complicated than that
of 1-instanton. In the limit of coincident instantons, which
attract the interests of many theoretical and mathematical
physicist recently and can be given in explicit form, the ADHM
data (\ref{Delta}) can be written as follows:
\begin{eqnarray}
& & \Delta=\left(\begin{array}{cccc} 0 &\sqrt{2\zeta}& 0 & 0\\
\bar z_2 & 0 & -z_1 & -\sqrt{\zeta}\\
0 & \bar z_2 & 0 & -z_1\\
\bar z_1 & 0 & z_2 & 0\\
\sqrt{\zeta}&\bar z_1 & 0 & z_2 \end{array}\right),
\\
& &  \Delta^\dag=\left(\begin{array}{ccccc} 0 & z_2 & 0 & z_1 &
\sqrt{\zeta}\\
\sqrt{2\zeta}& 0 & z_2 & 0 & z_1\\
0 & -\bar z_1 & 0 & \bar z_2 & 0\\
0 & -\sqrt{\zeta}& -\bar z_1 & 0 &\bar z_2 \end{array}\right).
\end{eqnarray}
By using this ADHM matrix, we have
\begin{equation}
f=\left(\begin{array}{cc} (Z_1''+Z_2'+\zeta)(Z''~')^{-1}
& -\sqrt{\zeta}(Z''~')^{-1}\bar z_1\\
-\sqrt{\zeta}z_1(Z''~')^{-1}& (Z_1'+Z_2'-\theta_1)(Z'~')^{-1}
\end{array}\right).
\end{equation}
where $Z''~'$ means $(Z_1''+Z_2')^2-\theta_1 Z_1''+\theta_2 Z_2'$,
$Z'~'$ means $(Z_1'+Z_2')^2-\theta_1 Z_1'+\theta_2 Z_2'$.

By following the same strategy as used in the 1-instanton case, we
find a general solution as follows:
\begin{equation}
U_0=\left(\begin{array}{c} (2\zeta)^{-1/2}Z\\
\sqrt{\zeta}\bar z_2\bar z_1\\
-\bar z_2(Z_1+Z_2+\theta_2)\\
\sqrt{\zeta}\bar z_1\bar z_1\\
-\bar z_1(Z_1+Z_2-\theta_1) \end{array}\right)
\end{equation}
where $Z=(Z_1+Z_2)^2-\theta_1 Z_1 +\theta_2 Z_2$. Now we discuss
the two $\theta_2$ cases separately.

\subsection{The $\theta_2>0$ case}

In this case $U_0$ annihilates two states: $|0,0\rangle$ and
$|1,0\rangle$ and so we must introduce one more shift operator
$\tilde{u}^\dag$ which satisfies the following relations:
\begin{equation}
\tilde u^\dag\tilde u=1-p_0-p_1,\quad\tilde u\tilde u^\dag=1,
\end{equation}
where $p_1=|1,0\rangle\langle 1,0|$. The correct $U$ is again
given by:
\begin{equation}
U=\tilde U_0\beta,\quad\tilde U_0=U_0\tilde u^\dag,
\end{equation}
where $\beta$ is a normalization factor:
\begin{equation}
\beta=(\tilde U_0^\dag\tilde U_0)^{-1/2}=(2\zeta)^{1/2}
(\tilde{u}ZZ'~'\tilde{u}^\dag)^{-1/2}.
\end{equation}

\subsection{The $\theta_2<0$ case}

There is a subtlety in this case. The matrix element
$f_{22}=(Z_1'+Z_2'-\theta_1)(Z'~')^{-1}$ of the matrix $f$ is not
well-defined on the vacuum $|0,0\rangle$. We remedy this
arbitrariness by the following definition:
\begin{equation}
f_{22}|0,0\rangle=(2\theta_1+\theta_2)^{-1}|0,0\rangle.
\end{equation}
With this definition one easily show that $f^{-1}f=ff^{-1}=1$.
What is $f^{-1}$.

In the $\theta_2<0$ case $U_0$ still annihilates $|0,0\rangle$.
Moreover, it turns out that we need three extra zero modes of
$\Delta^\dag$ to make $U_0$ complete. The matrix $U$ is
constructed as follow:

First, it is easy to find the following zero mode:
\begin{equation}
\Delta^\dag\left(\begin{array}{c} 0 \\ 0 \\ |0,0\rangle \\ 0 \\
0 \end{array}\right)=0
\end{equation}
is not included in $U_0$ and it is orthogonal to all columns of
$U_0$. So we can replace the redundant column of $U_0$ with this
extra zero mode and get
\begin{equation}
\tilde U_0=\left(\begin{array}{c} 0 \\ 0 \\ p_0 \\ 0 \\ 0
\end{array}\right)+U_0,
\end{equation}
which can be normalized as
\begin{equation}
\tilde{U}=\tilde U_0\beta,\quad\tilde{U}^\dag\tilde{U}=1
\end{equation}
where $\beta$ is a normalization factor:
\begin{equation}
\beta=(\tilde U_0^\dag\tilde U_0)^{-1/2}=(2\zeta)^{1/2}
(ZZ'~'+2\zeta p_0)^{-1/2}.
\end{equation}

Now we insert the other two (normalized) extra zero modes into
$\tilde{U}$ and get
\begin{equation}
U=\left(\begin{array}{c} \sqrt{\frac{-\theta_2}{2\theta_1+
\theta_2}}|0,0\rangle \\ 0 \\ -\sqrt{\frac{2\zeta}{2\theta_1+
\theta_2}}|0,1\rangle \\ 0 \\ 0 \end{array}\right)\langle 0,0|
+\left(\begin{array}{c} 0 \\ \sqrt{\frac{\theta_1}{2\theta_1+
\theta_2}}|0,0\rangle \\ -\sqrt{\frac{\zeta}{2\theta_1+\theta_2}}
|1,0\rangle \\ 0 \\ 0 \end{array}\right)\langle 1,0|
+\tilde{U}\tilde{u},
\end{equation}
which is the required $U$ and can be directly checked to satisfy
the completeness relation (\ref{complete}).

\section{$U(2)$ 1-instanton solution}

We will be brief here and rely on early results in the
$\theta_1=\theta_2$ case. The instanton positioned at the origin
and sized $\rho$ is determined from the following ADHM matrix:
\begin{equation}
\Delta=\left(\begin{array}{cc} \sqrt{\zeta+\rho^2} &
0 \\ 0 & \rho \\ \bar z_2 & -z_1 \\ \bar z_1 & z_2
\end{array}\right),
\quad\bar\Delta=\left(\begin{array}{cccc}
\sqrt{\zeta+\rho^2} & 0 & z_2 & z_1 \\
0 & \rho & -\bar z_1 & \bar z_2 \end{array}\right).
\end{equation}
The factorization relation is satisfied and we have
\begin{equation}
f=(Z_1+Z_2+\zeta+\rho^2)^{-1}.
\end{equation}

It is not difficult to find the general solution:
\begin{equation}
U_0=\left(\begin{array}{cc} Z_1+Z_2 & 0 \\ 0 & Z_1+Z_2+\zeta \\
-\sqrt{\zeta+\rho^2}\bar z_2 & \rho z_1 \\ -\sqrt{\zeta+\rho^2}
\bar z_1 & -\rho z_2 \end{array}\right)
\end{equation}
which should be  modified to lead to the correct $U$.

\subsection{The $\theta_2>0$ case}

It is easy to see that the first column of $U_0$ annihilates the
vacuum. So we can introduce an operator
\begin{equation}
v^\dag=\left(\begin{array}{cc} u^\dag & 0 \\ 0 & 1 \end{array}\right)
\end{equation}
to remove the redundance of $U_0$ and get the normalized $U$:
\begin{equation}
U=\tilde U_0\beta,\quad\tilde U_0=U_0 v^\dag
\end{equation}
where
\begin{equation}
\beta=(\tilde U_0^\dag \tilde U_0)^{-1/2}=\left(\begin{array}{cc}
[u(Z_1+Z_2)f^{-1}u^\dag]^{-1/2} & 0 \\ 0 &
[(Z_1+Z_2+\zeta)f^{-1}]^{-1/2} \end{array}\right).
\end{equation}
This matrix $U$ now satisfies the completeness relation
(\ref{complete}).

\subsection{The $\theta_2<0$ case}

As in the $U(1)$ theory, $U_0$ is directly normalizable:
\begin{equation}
\tilde{U}=U_0\beta,\quad\tilde{U}^\dag\tilde{U}=1
\end{equation}
where
\begin{equation}
\beta=(U_0^\dag U_0)^{-1/2}=\left(\begin{array}{cc}
[(Z_1+Z_2)f^{-1}]^{-1/2} & 0 \\ 0 & [(Z_1+Z_2+\zeta)
f^{-1}]^{-1/2} \end{array}\right).
\end{equation}
Again one more zero mode should be added to $\tilde{U}$ by using
the shift operator $v$ and the right $U$ is given by:
\begin{equation}
U=\left(\begin{array}{cc} 0 & 0 \\ 0 & 0 \\ p_0 & 0 \\ 0
& 0 \end{array}\right)+\tilde{U}v.
\end{equation}
One can again check that this $U$ satisfies the completeness
relation (\ref{complete}).

\section*{Acknowledgments}

Chuan-Jie Zhu would like to thank Prof. Zhu-Jun Zheng and the
hospitality at the Institute of Mathematics, Henan University.


\begin{thebibliography}{99}

\bibitem{WittenSeiberg} N. Seiberg and E. Witten, {\it String Theory
and Noncommutative Geometry}, J. High Energy Phys. 9909 (1999)
032, hep-th/9908142.

\bibitem{Seiberg} See, for example, S. Minwalla, M. van Raamsdonk
 and N. Seiberg,
{\it Noncommutative Perturbative Dynamics}, J. High Energy Phys.
0002 (2000) 020, hep-th/9912072; M. van Raamsdonk and N. Siberg,
{\it Comments on Noncommutative Perturbative Dynamics}, J. High
Energy Phys. 0003 (2000) 035, hep-th/0002186.

\bibitem{GSM} R. Gopakuma, S. Minwalla and A. Strominger,
{\it Noncommutative Solitons}, J. High Energy Phys. 0005 (2000)
020, hep-th/0003160.

\bibitem{Harvey} See, for example, J. A. Harvey, {\it Komaba Lectures
on Noncommutative Solitons and D-Branes}, hep-th/0102076.

\bibitem{Donaldson} S. K. Donaldson and P. B. Kronheimer, {\it The
Geometry of Four-Manifolds}, Clarendon Press, Oxford, 1990.

\bibitem{Schwarz} N. Nekrasov and A. Schwarz, {\it Instantons on
Noncommutative $R^4$, and (2,0) Superconformal Six Dimensional
Theory}, Commun. Math. Phys. 198 (1998) 689, hep-th/9802068.

\bibitem{Papera} K. Furuuchi, {\it Instantons on Noncommutative
$R^4$ and Projection Operators}, Prog. Theor. Phys. 103 (2000)
1043, hep-th/9912047.

\bibitem{Paperb} K. Kim, H. Lee and H. S. Yang, {\it Comments on
Instantons on Noncommutative $R^4$}, hep-th/0003093.

\bibitem{Paperc} N. Nekrasov, {\it Trieste Lectures on Solitons in
Noncommutative Gauge Theories}, hep-th/0011095.

\bibitem{Paperd} C.-S. Chu, V. V. Khoze and G. Travaglini, {\it
Notes on Noncommutative Instantons}, Nucl.Phys. B621 (2002) 101,
hep-th/0108007.

\bibitem{Reviewa} A. Konechny and A. Schwarz, {\it Introduction to
M(atrix) Theory and Noncommutative Geometry}, hep-th/0012145; {\it
Introduction to M(atrix) Theory and Noncommutative Geometry, Part
II}, hep-th/0107251.

\bibitem{Reviewb} M. R. Douglas and N. A. Nekrasov, {\it
Noncommutative Field Theory}, Rev. Mod. Phys. 73 (2002) 977,
hep-th/0106048.

\bibitem{Reviewc} R. J. Szabo, {\it Quantum Field Theory on
Noncommutative Spaces}, hep-th/0109162.

\bibitem{ADHM} M. F. Atiyah, N. J. Hitchin, V. G. Drinfeld and Y.
I. Manin, {\it Construction of Instantons}, Phys. Lett. A65 (1978)
185.

\bibitem{Nakajima} H. Nakajima, {\it Heisenberg Algebra and Hilbert
Schemes of Points on Projective Surfaces}, alg-geom/9507012 and
{\it Lectures on Hilbert Schemes of Points on Surfaces}.

\bibitem{Hamanaka} M.~Hamanaka,{\it ADHM/Nahm construction of localized solitons
in noncommutative gauge theories}, Phys.\ Rev.\ D {\bf 65} (2002)
085022, hep-th/0109070.

\end{thebibliography}
\end{document}